\documentstyle[twoside,fleqn,epsf,espcrc2]{article}
\input epsf.sty

\newcommand{\AmS}{{\protect\the\textfont2
  A\kern-.1667em\lower.5ex\hbox{M}\kern-.125emS}}

\begin{document}
\title{The Landau gauge lattice QCD simulation and the gluon propagator}

\author{Hideo Nakajima\address{Department of Information Science, 
        Utsunomiya University,\\
        2753 Ishii, Utsunomiya 321-8585, Japan 
       (e-mail nakajima@kinu.infor.utsunomiya-u.ac.jp)
        }%
        \thanks{Speaker at the conference.}
        and
        Sadataka Furui\address{School of Science and Engineering, 
        Teikyo University, \\
         1-1 Toyosatodai, Utsunomiya 320-8551, Japan
         (e-mail furui@dream.ics.teikyo-u.ac.jp)}}

\begin{abstract}
The gluon propagator in the Landau gauge lattice QCD simulation is
measured. The data suggests the confinement mechanism of the Gribov-Zwanziger theory.

\end{abstract}

\maketitle

\section{Introduction}
 In 1978 Gribov showed that the fixing of the divergence of the gauge
field in non-Abelian gauge theory does not fix its gauge\cite{Gv}.
The restriction of the gauge field such that its Faddeev-Popov
determinant is positive (Gribov region) is necessary yet insufficient
for excluding the ambiguity. 

 The restriction to the fundamental modular region\cite{Zw,lat98} cancels the 
infrared singularity of the perturbation theory and makes the gluon to have the
 complex mass. 
The transverse gluon propagator in the continuum theory is defined as

\begin{eqnarray}
D_{\mu\nu}(k)&=&\displaystyle{{1\over n}}
\sum_{x={\bf x},t} e^{-ikx}Tr\langle A_\mu(x) A_\nu(0) \rangle\nonumber\\
&=&G(k^2)(\delta_{\mu\nu}-{k_\mu k_\nu\over k^2})\ \ .
\end{eqnarray}
where $n$ is the dimension of the adjoint representation of SU(N). 
Schematically
\begin{equation}
 G(k^2)={k^2\over k^4+\kappa^4}={1\over 2}({1\over k^2+i\kappa^2}+
{1\over k^2-i\kappa^2}).
\end{equation}

It was shown\cite{Zw} that in the lattice simulation of the
non-Abelian gauge theory, the Landau gauge fixing and the restriction
of the gauge field yields the gluon propagator similar to that of Gribov.

\section{The global gauge fixing and the gluon propagator} 

We performed  quenched ($4^3\times 8$ and $8^3\times 16$) and unquenched 
($4^3\times 8$) lattice QCD simulation in the Landau gauge whose algorithm
is presented in\cite{lat97,lat98}. The number of configuration is 100 for the
three cases. 

After the Landau gauge fixing with an accuracy ${\rm Max}\vert \partial_\mu A_\mu
\vert<10^{-4}$ which is roughly equivalent to $(\partial_\mu A_\mu)^2<
10^{-10}$, we fix the global gauge of the field $A_\mu(x)$ such that 
$A=\sum_{\mu x a} A^a_\mu(x)\lambda^a$ of a sample is diagonalized by an
SU(3) matrix $g$ : $g^\dagger A g=diag(g_1,g_2,g_3),(g_1\ge g_2\ge g_3)$.  

We obtain the new data sets ${A^g}_\mu(x)=g^\dagger A_\mu(x) g$
and perform the Fourier transform
$A^{ga}_\mu(n_k)={1\over \sqrt {N_x^3N_t}}\sum_x e^{-ik\cdot (x+e_\mu/2)}A^{ga}_\mu(x)$.

Using the projection operator on a lattice in the momentum space
\begin{equation}
P_{\mu\nu}(n_k)=\delta_{\mu\nu}-{c_\mu(n_k) c_\nu(n_k) \over c(n_k)^2}
\end{equation}
where $c_\mu(n_k)=2{\rm sin} {\hat k_\mu\over 2}$, 
$c(n_k)^2=\sum_{\mu=1}^4 c_\mu(n_k)^2$
and $\hat k_\mu={2\pi n_k\over N_\mu}$ ($n_k=0,\cdots ,N_\mu -1$) is the 
available momentum on the
lattice in the $\mu-$direction, the gluon propagator is calculated from
\begin{eqnarray}
D(\hat k)&=&\sum_{\mu\nu a} P_{\mu\nu}(n_k)(\langle {A^{g a}}_\mu(n_k) 
{A^{g a}}_\nu(n_k)^*\rangle\nonumber\\
&-&\langle {A^{g a}}_\mu(n_k)\rangle\langle {A^{g a}}_\nu(n_k)^*\rangle)
\end{eqnarray}
where $\langle {A^{g a}}_\mu(n_k)\rangle$ is the average over samples of
the globally gauge fixed field $A^{g a}_\mu(n_k)$. 
 
We define the gluon propagator as a function of
$\hat k=\sqrt{ \sum_\mu \hat k_\mu^2}$ where $\hat k_\mu={\rm Min}[n_k, N_\mu-n_k]
{2\pi\over N_\mu}$ and not a function of
 $k=\sqrt{ \sum_\mu (2 {\rm sin}{\hat k_\mu\over 2})^2}$. 

In the case of $8^3\times 16$ lattice there appear 4096 independent 
four-momentum $\hat k_\mu$.
For a fixed $\hat k$, the data of $D(\hat k)$ from $0\le n_k < {N_\mu\over 2}$
and ${N_\mu\over 2} \le n_k' <N_\mu$  which satisfies $N_\mu-
n_k'=n_k$ are plotted at the same abscissa and in the medium momentum region
the difference of the data at $n_k'$ and $n_k$ makes the plot of
$D(\hat k)$ scattered. This effect could be attributed to the finite size
effect and in larger lattice the effect becomes smaller\cite{LSWP}.

 We adopt the data selection prescription 
similar to that of \cite{LSWP}, i.e. select data such that 
$\vert n_{ki}-n_{kj}\vert\le 1$ where $i$ and $j$ run from 1 to 4, or choose 
the momentum $\hat k$ such that
it is almost diagonal in the $N_x^3 {N_t\over 2}$ lattice space.
In fig. 1 we present the result of $D(\hat k)$. The data clearly shows
the infrared suppression of the gluon propagator. The suppression remains
even without the subtraction of $\langle {A^{g a}}_\mu(n_k)\rangle\langle
 {A^{g a}}_\nu(n_k)^*\rangle$ term.

We performed the analysis also for $\hat k$ along the spacial 
axis\cite{conf3}.  The results of
$8^3\times 16$ lattice suggests that the rotational symmetry is
not so good. Although in the quenched $4^3\times 8$ lattice  
the gluon propagator did not show the significant suppression and remained
finite, in the unquenched $4^3\times 8$ lattice  similar
suppression was observed\cite{conf3}.    
In the SU(2) case similar effect is also observed\cite{Cu}.

\begin{figure}[hbt]
\begin{center}
\leavevmode
\epsfysize=150pt\epsfbox{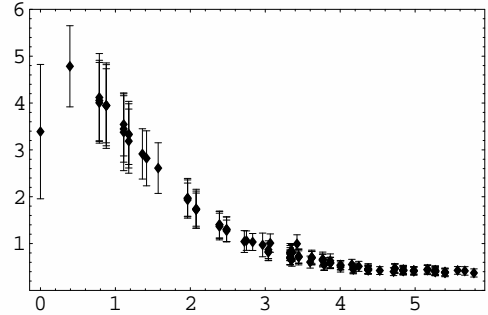}
\end{center}
\caption{The gluon propagator $D(\hat k)$ of quenched $\beta=5, 8^3\times 16$
lattice.}
\end{figure}

\section{Parametrization of the gluon propagator}

In order to get insights on the complex mass of the confined gluon, 
we performed an analysis of the Fourier transform of the correlation function
\begin{equation}
D_T({\bf k},t)={1\over 2n}\sum_{jTx}e^{i{\bf k}\cdot{\bf x}}
Tr\langle A_j({\bf x},T+t)A_j({\bf 0},T)\rangle_c
\end{equation}
where the suffix c indicates that the connected part is taken after the
global gauge fixing and
\begin{equation}
\tilde A^a_\mu({\bf k},T)={1\over N_x^3}\sum_x A^a_\mu({\bf x},T)e^{-i{\bf k}\cdot{\bf x}}.
\end{equation}

In the Stingl's factorizing-denominator rational approximants
(FDRA) method\cite{St}, the transverse gluon propagator is expressed as
\begin{equation}
D_T(p^2)^{[r=3]}={\rho\over p^2+u_+\Lambda^2}+{\tau\over p^2+v_+\Lambda^2}+c.c.
\end{equation}
which is compared with its lattice Fourier transform
\begin{eqnarray}
&&G_T({\bf k},t)_c=Re[(a_0+ i a_1){\rm cosh}((a_2+i a_3)(t-4))\nonumber\\
                &+&(a_4+ i a_5){\rm cosh}((a_6+i a_7)(t-4))]\nonumber\\
&=&Re[{\rho a\over {\rm sinh}( \hat M_1a) {\rm sinh}({N_T\over 2}\hat M_1a)}
              {\rm cosh}(\hat M_1a (t-4))\nonumber\\
 &+&{\tau a\over{\rm sinh} (\hat M_2a){\rm sinh}({N_T\over 2}\hat M_2a)}
               {\rm cosh}(\hat M_2a (t-4))]
\end{eqnarray}
where the gluon masses are defined as $\hat M_1a=a_2+i a_3, \hat M_2a=a_6+i a_7$ 
and
\begin{equation}
\rho a=(a_0+i a_1) {\rm sinh}(a_2+i a_3) {\rm sinh}(N_T (a_2+i a_3)/2),
\end{equation}
\begin{equation}
\tau a=(a_4+i a_5) {\rm sinh}(a_6+i a_7) {\rm sinh}(N_T (a_6+i a_7)/2),
\end{equation}
\begin{equation}
u_+\Lambda^2=4 {\rm sinh}^2((a_2+i a_3)/2),
\end{equation}
\begin{equation}
v_+\Lambda^2=4 {\rm sinh}^2((a_6+i a_7)/2).
\end{equation}

\begin{figure}
\leavevmode
\begin{center}
\epsfysize=150pt\epsfbox{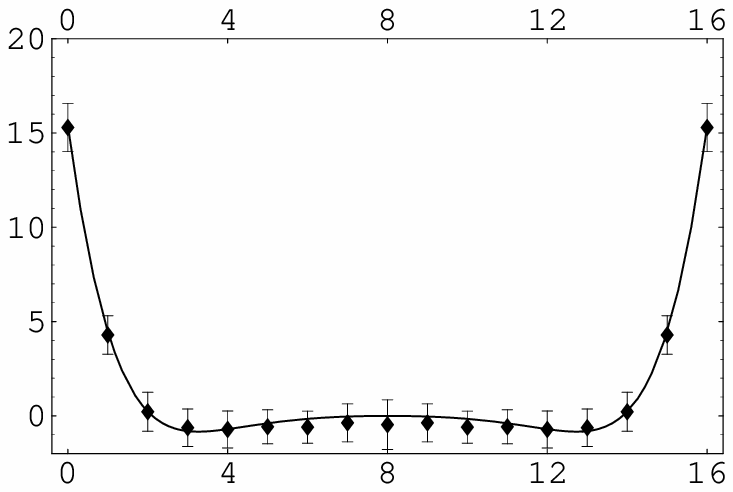}
\end{center}
\caption{The gluon propagator $G_T({\bf 0},t)$ of quenched $\beta=5, 8^3\times 16$ lattice.
}
\begin{center}
\epsfysize=150pt\epsfbox{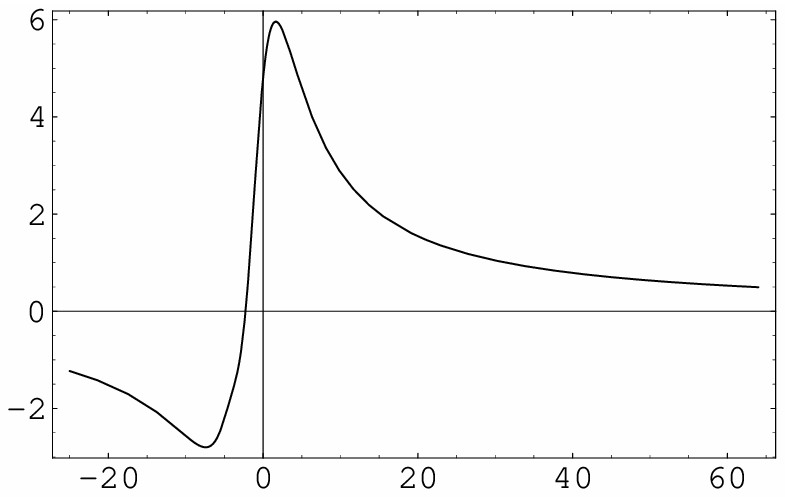}
\end{center}
\caption{The lattice Fourier transformed gluon propagator $D(p^2)$.}
\end{figure}
 
The gluon propagator $D(p^2)$ derived in the FDRA method also shows infrared
suppression. We observe that as the lattice size becomes large, the position 
of zero of $D(p^2)$ approaches to $p^2=0$ and that the
peak near $p^2=0$ becomes sharp.

\section{Discussion and outlook}

 We performed the Langevin simulation of the quenched and unquenched 
lattice QCD in the Landau gauge, using the natural definition of the gauge
field.

The position of the zero of the Fourier transform of the connected part of the
 gluon
propagator  $D_T(p^2)$ approaches $p^2=0$ as the lattice size increases, which
suggests the possibility of the realization of the confinement mechanism of the
 Gribov-Zwanziger theory.

Our preliminary fitting results indicate that real part of the effective 
mass of the gluon is about $1.4a^{-1}$ for $\beta=4, 4^3\times 8$ 
and about $0.75a^{-1}$ for $\beta=5, 8^3\times 16$. 
The renormalization and the anomalous dimension for the effective mass of the
gluon is under investigation.

This work was supported by High Energy Accelerator Research Organization, as
KEK Supercomputer Project(Project No.98-34).


\begin{thebibliography}{20}
\bibitem{Gv} V.N. Gribov, Nucl. Phys. {\bf B139} (1978) 1.
\bibitem{Zw} D. Zwanziger, Nucl. Phys. {\bf B364} (1991) 127.
\bibitem{lat97} H. Nakajima and S. Furui, Nucl Phys. {\bf B} (Proc. Suppl.)
 63 (1998) 976, hep-lat/9710028.
\bibitem{lat98} H. Nakajima and S. Furui, in this proceedings. 
\bibitem{LSWP} D.B. Leinweber, J.I Skullerud, A.G. Williams and C. Parrinello,
Phys. Rev.{\bf D58} 031501, hep-lat/9803015
\bibitem{conf3} H. Nakajima and S. Furui, Confinement III proceedings, 
June 1998, Jefferson Lab, NewPort.  
\bibitem{Cu} A. Cucchieri, Phys. Lett. {\bf B422} (1998) 233, hep-lat/9709015. 
\bibitem{St} M. Stingl, Z. Phys. {\bf A353} (1996) 423, hep-th/9502157;
J. Fromm and M. Stingl, preprint, M\"unster University(1997)
\end{thebibliography}
\end{document}